# Impact of the Overall Electrical Filter Shaping in Next-Generation 25G and 50G PON

Pablo Torres-Ferrera[1*], Valter Ferrero[1], Maurizio Valvo[2], and Roberto Gaudino[1]

[1] Politecnico di Torino, Dipartimento di Elettronica e Telecomunicazioni, Torino, Italy. * pablo.torres@polito.it.
[2] M. Valvo is with Telecom Italia Mobile (TIM), Torino, Italy.

*Abstract*—Next-generation high-speed passive optical network (HS-PON) transceivers supporting 25, 50 and 100 Gb/s are under the early stage of their standardization process. One key aspect of this process is the choice of the best modulation format. To this end, performance comparisons among several modulation formats against different physical constraints have been presented in literature and are still being carried out. In our contribution, we performed an exhaustive analysis on the impact of transceivers electrical frequency response shape on the performance of 2-levels pulse amplitude modulation (PAM-2), 4-levels PAM (PAM-4), electrical and optical duobinary modulation formats with adaptive equalizer at the receiver side. We show by means of numerical simulations that the specification of the typically used -3dB bandwidth is not sufficient, since also out-of-band electrical frequency response specifications (such as the -20dB bandwidth) has a huge impact on the performance of the analyzed modulation formats. The normalized graphs given at the end of the paper in terms of -3dB and -20 dB bandwidths can thus be useful for the design of the next generation of HS-PON transceivers.

1. INTRODUCTION

Standardization efforts to define the physical layer characteristics of next-generation high-speed passive optical networks (HS-PON) are currently being carried out in the International Telecommunications Unit (ITU-T), Full Service Access Network (FSAN) and Institute of Electrical and Electronics Engineers (IEEE) standardization bodies [1–5]. Several research analyses are currently ongoing to choose the best modulation format for HS-PON transceivers for the different bit rate under consideration (such as 25, 40 and 50 Gb/s) [6-17]. Due to the PON low cost constraint, particularly on the optical network unit (ONU) (i.e. user) side, several groups are considering if the transmitter and receiver optoelectronics developed for lower bit rates can be re-used for the new higher bit rates when associated to more bandwidth efficient modulation formats and/or adaptive equalization. In particular, it would be interesting to re-use the optoelectronics developed for 10 Gb/s PON also for the 25 Gb/s target (or even the 50 Gb/s one), and similarly the 25 Gb/s transceivers developed for the intra-datacenter short-range transceiver also for 40-50 Gb/s PON.

The target of this work is thus to analyze the feasibility of these goals, focusing on the fact that these transceivers would be severely electrical bandlimited when used for the new target bit rates envisioned for HS-PON (25 and 50 Gb/s). The novelty of this paper is to show the importance of the complete frequency response of the available optoelectronics. While previous papers in this area focus usually on the -3dB bandwidth only [18], or experimentally on a single given transceiver, our added value is to extensively study the impact of different overall frequency transfer functions on system performance. We show that the "out-of-band" frequency response is quite important for the modulation formats under consideration, that are PAM-2 (i.e. binary On-Off Keying), PAM-4, electrical duobinary (EDB) [6, 11] and optical duobinary (ODB) [8] and when using traditional direct-detection (DD) receivers followed by digital signal processing (DSP)-based adaptive equalizer. Specifically, we give normalized graphs showing the joint impact of the -3dB and -20 dB parameters of the available frequency response. We believe that these graphs can give a useful contribution to the discussion currently ongoing in the aforementioned HS-PON standardization bodies and to the transceiver vendors that, depending on the details of their optoelectronic technology, can better select how to optimize their component design.

Thanks to our simulative approach, we were able to span a very large set of parameters (-3dB bandwidth, -20 dB bandwidth, accumulated dispersion, etc.) to obtain power penalty curves for specific bit error rate (BER) values and to extensively compare the four different modulation formats. Moreover, we superimposed to our simulative results the expected bandwidths of several existing transmitter and receivers, thus giving a very broad review of the existing literature in this field.

Summarizing our previous considerations, we believe that the main novelty of this paper is in giving "design rules" for the transceivers full frequency response shapes for different modulation formats. Moreover, our analysis points out the different resilience of each modulation formats to variations in the available bandwidth, an information that again can be interesting for component designers.

To this end, the paper is organized as follows. In Section 2, we present the details of the considered transceiver and link architecture and of our simulation environments. In Section 3, we discuss on different frequency response shape, and we apply it in Section 4 evaluating the resulting power budget penalty vs. filter shape parameters. In the following Section 5 we also introduce the impact of the typical chromatic dispersion accumulated in PON for different choice of the wavelength band. Finally, we draw the conclusion in Section 6.

## 2. SIMULATION SETUP

The transceiver and link architecture that we considered in our simulation setup is shown in Fig. 1, where the variable optical attenuator (VOA) will be used to span different values of link loss and, thus, the impact of the 1xN PON splitter. For space limitation we will not consider penalty arising from burst mode transmission. Apart from this (important [19]) consideration, our analysis can be applied to both downstream and upstream directions of a PON link.

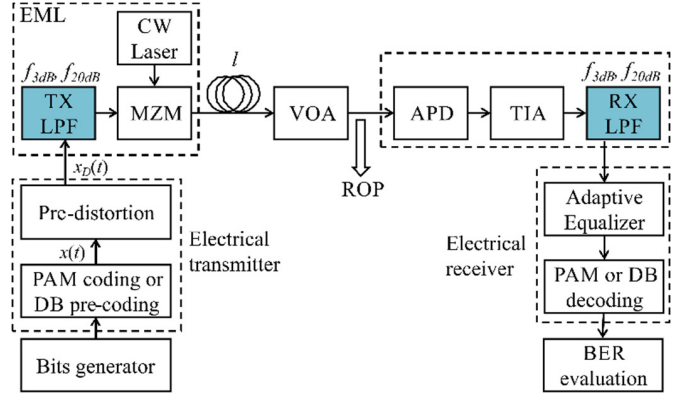

Fig. 1. Simulation setup.

At the transmitter side, a binary signal at bit rate $R_b$ (equal to either 25 or 50 Gb/s) is generated by means of a pseudo-random bit sequence (PRBS) $2^{17}$ -1 bits long. The bit stream feeds the Electrical TX that creates the appropriate driving signal to generate the desired: PAM-2, PAM-4, EDB or ODB signals in the optical domain. We considered only external modulation in this paper (or more in general optical transmitters characterized by negligible chirp). The generated electrical signal indicated as $x(t)$ in Fig. 1 is obtained in different ways depending on modulation formats:

i) In the PAM-2 and PAM-4 cases, the binary signal is mapped into a 2-levels (-1 and +1, also called antipodal binary signal) or 4-levels (-3, -1, +1, and +3) symbol stream, respectively. Gray coding is used in PAM-4.

ii) In both EDB and ODB cases, the binary signal is pre-coded by applying a standard XOR-based scheme [11]. The 2-levels pre-coded signal (called here "pre-DB") will turn into a 3-levels DB signal thanks to the intrinsic low-pass filtering response of the transceiver (thus avoiding the use of additional add-and-delay encoding or low-pass electrical filter circuits).

Time-domain simulation are performed using 8 samples per bit (*spb*). The signal $x(t)$ is normalized to $0 \leq x(t) \leq 1$. We than assume that the DSP can compensate the Mach-Zehnder optical modulator (MZM) non-linear response by applying the following pre-distortion:

$$x_D(t) = \frac{A}{\pi} \arccos\left(1 - 2x(t)\right) - V_b. \tag{1}$$

where $A$ is an amplitude factor and $V_b$ is the bias-voltage of the MZM. The pre-distorted signal $x_D(t)$ is then filtered using an electrical low pass filter (LPF) that emulates the electrical frequency response of the transmitter (TX). The shape of this filter (and the following one at the receiver) is one of the main goal of our investigation. Details on the assumed filter shapes will be given in the following Sect. II. The resulting electrical signal after filtering $x_F(t)$, drives an optical modulator, optically fed by a continuous wave (CW) electrical field, $E_{CW}(t)$, generated by the TX laser. The electrical LPF, the optical modulator and the laser composes the externally modulated laser (EML) block. The optical signal at the output of the EML is modeled using a classical (chirp-less) MZM equation:

$$E_{EML}(t) = E_{CW}(t) \cos \frac{\pi x_F(t)}{V_\pi}, \tag{2}$$

where $V_\pi$ is the π-voltage of the modulator. By setting both parameters $A$ and $V_b$ of Eq. (1) equal to $V_\pi/2$ for PAM-2, PAM-4 and EDB, or equal to $V_\pi$ for ODB, the MZM is operated in quadrature or null, respectively. The modulated optical signal is then propagated through a conventional single mode fiber (SMF). Only chromatic dispersion is considered in the fiber model since the non-linear effects are assumed negligible for the relatively short reach (≤20km) applications under study. The received optical signal is detected by means of an avalanche photodetector (APD) at the receiver (RX) side followed by a trans-impedance amplifier (TIA). APD+TIA configurations seems to be the most likely to be applied for 40-Gigabit-capable PON (NG-PON2) and 10-Gigabit-capable symmetric PON (XGS-PON), and this is why we decided to focus on them. The photocurrent that outputs the APD+TIA is evaluated by the following expression:

$$i(t) = GRP(t) + n_S(t) + n_T(t), \tag{3}$$

where $R$ is the APD responsivity (assumed to be $R = 0.8$ A/W), $G$ is the APD gain factor (assumed to be $G = 25$ or 14 dB) and $P(t)$ is the optical signal instantaneous power that feeds the APD. The signals $n_S(t)$ and $n_T(t)$ emulate the shot noise and thermal noise, respectively. They are modeled as Gaussian random process with zero-mean and variances given by [20]:

$$\sigma_S^2(t) = qG^2 FRP(t)\Delta f_s, \qquad (4)$$

$$\sigma_T^2 = N_0 \Delta f_s, \qquad (5)$$

respectively, where $F$ is the APD excess noise factor (we assumed $F = G^{0.75} = 10.5$ dB), $q$ is the electron charge, $N_0$ is the input-referred electrical current thermal noise power spectral density ($N_0 = 1.024 \times 10^{-21}$ A$^2$/Hz), and $\Delta f_s$ is the bandwidth of the simulation ($\Delta f_s = spb \cdot R_b$). The overall thermal noise of APD and TIA contributions are included into the $n_T(t)$ noise term. An electrical LPF is placed after the APD to emulate the RX frequency response. The characteristics of this RX LPF are the same as the TX LPF, and they are described in next Section 3. We know that this is not the more general case since typically the receiver filter shape is independent on the transmitter filter shape. Anyway, we introduce this assumption to limit the number of free parameters. In fact, we will show in Section 3 that we will consider two degrees of freedom for each electrical filter (for instance the -3dB and the -20dB bandwidth). Introducing the assumption of identical TX and RX filter we thus have 2 degrees of freedom, which we can represents using 2D contour plots, while assuming independent TX and RX filter we would have 4 degrees of freedom to span, which would be very difficult even just to be graphically presented, not to talk about the required simulation time.

After the RX LPF, since we want to focus on severely bandlimited situations, the received electrical signal is equalized by means of an adaptive direct feed-forward equalizer (FFE) using least-mean square (LMS) as adaptation algorithm with two samples per symbol and 20 taps. The equalizer is trained with a pilot sequence: in the case of PAM-2, and PAM-4, a 2-level and 4-level symbol stream, respectively; in the case of EDB, a 3-level DB symbol stream obtained after encoding a 2-level pre-DB signal by means of and add-and-delay block; and in the case of ODB a bit (antipodal) stream. The equalized signal is then decoded according to the employed modulation format. Finally, the bit error rate (BER) is evaluated by means of direct error counting over $1.3 \times 10^5$ bits (after the training sequence), a situation that is very CPU-time demanding but allows very precise estimation of target BER around $10^{-3}$.

The main figure of merit employed in this work to evaluate the performance of the system is the sensitivity (S), defined as the received optical power (ROP) in dBm to reach a BER target of $10^{-3}$. The ROP is measured at the APD input as the mean value over time of $P(t)$. A VOA is placed before the APD in order to sweep the ROP parameter.

## 3. FILTERING CONSIDERATIONS

We emulate these "narrow band" transceivers by using two LPF electrical filters, one at the transmitter side and one at the receiver side. One key parameter of a transceiver frequency response is obviously its 3-dB electrical bandwidth. Most of the experimental and commercial devices characterizations provide information about this parameter. However, as the main target of our work, we focus on the fact that not only the $f_{3dB}$ parameter is fundamental, but also the out-of-band (i.e. $f > f_{3dB}$) frequency response of the devices has a huge impact on the overall system performance. For instance, we show that for the 25 Gb/s target, $f_{3dB} = 8$ GHz can have very different penalties depending on the actual value of, say, the -20dB bandwidth. Moreover, we will show that the sensibility versus this parameter is very different depending on the used modulation format. To the best of our knowledge, this is a novel analysis for the scenario of HS-PON, since the out-of-band transceiver electrical frequency characterization is barely ever considered in details in the available literature. To investigate the relevance of this out-of-band filter shaping, we introduce the 20-dB bandwidth parameter ($f_{20dB}$) in our present analysis. The joint impact of the $f_{3dB}$ and the $f_{20dB}$ parameters is then tested, providing an extra degree of information regarding the filtering shape impact on the performance.

Among all the LPF options to emulate the frequency response of the transceivers, the Butterworth (BF) and Super-Gaussian filter (GF) profiles were chosen. The former was selected since it is straightforward to exactly set the desired $f_{3dB}$ parameter and to characterize it in terms of number of poles. However, in BF and for a fixed $f_{3dB}$, the value of the $f_{20dB}$ parameter is subject to the choice of the number of poles of the filter, that is an integer number and, consequently, $f_{20dB}$ can be only changed over given (and quite coarse) discrete values. Therefore, for BF it is not possible to set any arbitrary combination of the $f_{3dB}$ and $f_{20dB}$ parameters. The use of GFs allows overcoming this limitation by allowing an independent variation of $f_{3dB}$ and $f_{20dB}$. The frequency profile of a GF is defined here as:

$$H(f) = \exp\left(-\frac{1}{2}\left(\frac{f}{f_0}\right)^{2n}\right), \qquad (6)$$

where $n$ is the order of the GF filter (which now does not need to be necessarily and integer) and $f_0$ is a free parameter. The two free real values $n$ and $f_0$ can be set to obtain a frequency response having any possible combination of $f_{3dB}$ and $f_{20dB}$ (as long as $f_{20dB} \geq f_{3dB}$). In Fig. 2 the filter shape profiles of the a) BF and b) GF models are presented for different number of poles and order, respectively. The frequency responses of a TX and a RX reported in literature [17] are also displayed in the figure. Since they have a different $f_{3dB}$, the frequency values of the TX1 EML and RX1 APD curves in Fig. 2 were normalized with respect to its own $f_{3dB}$ value for graphical purposes, to perform a comparison in terms of filter shaping with respect to the BF and GF filter profiles analyzed. From Fig. 2 it can be seen that the GF model allows improving the accuracy of the out-of-band frequency response emulation, since the $f_{20dB}$ can be set as a continuous parameter, in contrast with the BF approach in which the $f_{20dB}$ is a discrete parameter subject to the choice of the order of the filter, as mentioned before. We anyway decided to keep using also the BF approach in the rest of this document for comparison purposes.

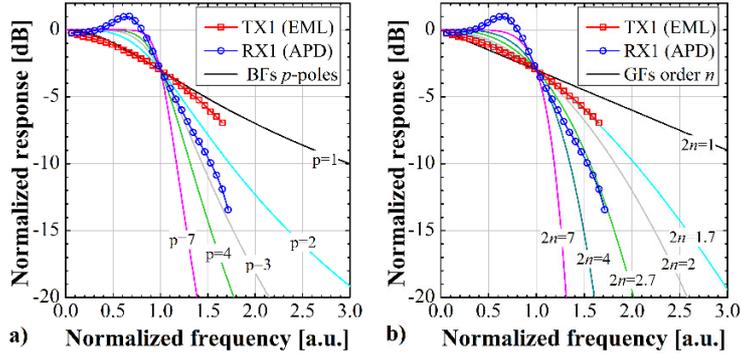

Fig. 2. Butterworth (BF) and Gaussian filter (GF) profiles, for different number of poles and order, respectively. The normalized experimental response of the TX (EML) and the RX (APD) found in literature [ref] are also shown.

The most significant assumption made in this work is that the TX and RX LPFs have exactly the same filter shape characteristics. Although, this approach may appear not very realistic for the experimental implementation point of view, a systematic analysis testing different combinations of $f_{3dB}$ and $f_{20dB}$ values in both the TX and RX filters would require studying a huge number of cases and several degrees of freedom, which in turn results impractical. Therefore, we preferred to make the aforementioned assumption, since anyway the equivalent filter formed after cascading any pair of TX and RX LPFs having the same characteristics could approach the overall equivalent frequency response formed by any particular combination of real TX and RX devices with different filter shaping, so TX and RX with different $f_{3dB}$ and $f_{20dB}$ values, as exemplified in Fig. 3. The frequency response of the devices shown in Fig. 3 are the same represented in Fig. 2 (TX1 EML $f_{3dB}$ = 8.9 GHz, RX1 APD $f_{3dB}$ = 7.5 GHz). The frequency response that results after cascading these TX1 EML and RX1 APD devices is shown in the solid black curve. Two identical GFs (or BFs), so both with the same $f_{3dB}$ and $f_{20dB}$, one at the TX and one at the RX, are cascaded and their combined frequency response is shown in the green dashed (or pink dotted for a pair of identical BFs) curve. In the case of GFs, the best matching with the solid black curve (the real devices with different filter shaping) was obtained with two identical GFs with $f_{3dB}$ equal to 8.5 GHz and $f_{20dB}$ equal to 17 GHz. The equivalent pair of identical BFs are 2-poles filters with $f_{3dB}$ equal to 8.1 GHz. By following this same procedure, we evaluated the $f_{3dB}$ and $f_{20dB}$ values of the GFs whose cascaded responses best matched the cascaded response of several TX and RX devices reported in the literature (both commercial and experimental). The results are summarized in Table 1. Please note that all these TX and RX devices were developed for the 10 Gb/s transmission, then we label them using "10G technology". In Table 2, the same information provided by Table 1 is reported but now corresponding to TX and RX devices reported in literature for the 25 Gb/s transmission of, called here "25G technology". In both tables, the parameters $B_{3dB}$ and $B_{20dB}$ are a normalized representation of $f_{3dB}$ and $f_{20dB}$, respectively, expressed in percentages and defined as:

$$B_{3dB}[\%] = \frac{f_{3dB}}{R_b} \times 100 \text{, and } B_{20dB}[\%] = \frac{f_{20dB}}{R_b} \times 100 \cdot \qquad (7)$$

So the same couple of TX and RX devices developed for 10G technology, if used in the 25 Gb/s transmission are equivalent of two identical GFs shaping with $B_{3dB}$ and $B_{20dB}$ parameters. If we use the same TX and RX devices but in the 50 Gb/s transmission the equivalent couple of GFs is the same but has different $B_{3dB}$ and $B_{20dB}$ values with respect to the previous 25 Gb/s case. The changing in the $B_{3dB}$ and $B_{20dB}$ parameters is due to the performed normalization, but the equivalent couple of GFs does not change in both cases.

$B_{3dB}$ and $B_{20dB}$ are used to present most of the results in the rest of this document. Information provided in Table 1 and Table 2 is helpful to contextualize our results in the framework of state-of-the-art technology.

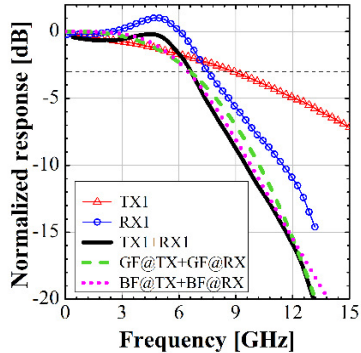

Fig. 3. Butterworth and Gaussian filters profile

Table 1. Normalized 3-dB and 20-dB bandwidth of the GF at TX and RX ($B_{3dB}$ and $B_{20dB}$) equivalent to the TX + RX 10G components cascade, reported in the references to be used in 25 or 50 Gb/s transmission

| C. | TX | | RX | | If $R_b$ = 25 | | If $R_b$ = 50 | |
|---|---|---|---|---|---|---|---|---|
| | Ref | $f_{3dB}$ | Ref | $f_{3dB}$ | $B_{3dB}$ | $B_{20dB}$ | $B_{3dB}$ | $B_{20dB}$ |
| 1 | [17] | 8.9 | [17] | 7.5 | 34 | 68 | 17 | 34 |
| 2 | [17] | 8.9 | [22] | 8.8 | 32.8 | 112 | 16.4 | 56 |
| 3 | [17] | 8.9 | [23] | 8.1 | 32 | 136 | 16 | 68 |
| 4 | [17] | 8.9 | [24] | 6.8 | 29.6 | 96 | 14.8 | 48 |
| 5 | [21] | 7.7 | [17] | 7.5 | 32 | 56 | 16 | 28 |
| 6 | [21] | 7.7 | [22] | 8.8 | 27.6 | 80 | 13.8 | 40 |
| 7 | [21] | 7.7 | [23] | 8.1 | 28 | 96 | 14 | 48 |
| 8 | [21] | 7.7 | [24] | 6.8 | 26.8 | 72 | 13.4 | 36 |
| 9 | * | 9.9 | [17] | 7.5 | 34 | 70 | 17 | 35 |
| 10 | * | 9.9 | [22] | 8.8 | 35.2 | 100 | 17.6 | 50 |
| 11 | * | 9.9 | [23] | 8.1 | 34.8 | 128 | 17.4 | 64 |
| 12 | * | 9.9 | [24] | 7.5 | 31.6 | 92 | 15.8 | 46 |
| 13 | TX + RX [25] | | | | 33.2 | 70 | 16.6 | 35 |

C.: Case. $f_{3dB}$, in GHz. $B_{3dB}$ and $B_{20dB}$ in %. $R_b$ in Gb/s.
* NG-PON2 TX characterization provided by TIM.

Table 2. Normalized 3-dB and 20-dB bandwidth of the GF at TX and RX ($B_{3dB}$ and $B_{20dB}$) whose cascade response best approaches the cascade response of the TX + RX 25G components reported in the references.

| C. | TX | | RX | | $R_b$ = 50 Gb/s | |
|---|---|---|---|---|---|---|
| | Ref | $f_{3dB}$ [GHz] | Ref | $f_{3dB}$ [GHz] | $B_{3dB\%}$ | $B_{20dB\%}$ |
| 1 | [26] | 18.9 | [28] | 15.8 | 34 | 60 |
| 2 | [26] | 18.9 | [29] | 24.5 | 40 | 132 |
| 3 | [26] | 18.9 | [30] | 32 | 45 | 140 |
| 4 | [26] | 18.9 | [31] | 19.9 | 40 | 88 |
| 5 | [27] | 28.2 | [28] | 15.8 | 32 | 58 |
| 6 | [27] | 28.2 | [29] | 24.5 | 50 | 130 |
| 7 | [27] | 28.2 | [30] | 32 | 50 | 110 |
| 8 | [27] | 28.2 | [31] | 19.9 | 44 | 88 |
| 9 | [21] | 23.9 | [28] | 15.8 | 30 | 58 |
| 10 | [21] | 23.9 | [29] | 24.5 | 42 | 84 |
| 11 | [21] | 23.9 | [30] | 32 | 44 | 90 |
| 12 | [21] | 23.9 | [31] | 19.9 | 38 | 78 |

## 4. BACK-TO-BACK RESULTS

A back-to-back (BtB) performance comparison among the four modulation formats when varying the $B_{3dB}$ parameter was first performed. As a first approach, 1-pole BFs were considered to emulate the electrical frequency response of each TX and RX device. In the inset of Fig. 4.a. are depicted the filter profiles of the individual TX (or RX) 1-pole BF (solid), and the cascaded response of them (dotted), for a $f_{3dB}$ = 7 GHz (single filtering). To enable a fair comparison, the performance is evaluated in terms of power penalty with respect to the sensitivity of PAM-2 in the best condition ( without any BW limitations and in a BtB scenario). This sensitivity (the ROP to guarantee $10^{-3}$ BER) is termed S0, and it is equal to S0 = -28.1 dBm for $R_b$ = 25 Gb/s, and S0 = -25.7 dBm for $R_b$ = 50 Gb/s. The computed power penalty of each modulation format as a function of the $B_{3dB}$ parameter is shown in Fig. 4.a, for both analyzed bit rates: 25 Gb/s (solid) and 50 Gb/s (dotted). A very close agreement between the results of the two analyzed bit rates was found (the difference between the solid and dotted curves is negligible irrespective of the modulation format). By using the information provided in Table 1, a pair of colored regions that show the range of $B_{3dB}$ values that current 10G technology transceiver exhibits to transmit 25 Gb/s (in blue) and 50 Gb/s (in yellow) are also depicted in Fig. 4.

In order to explore the impact of the out-of-band filter shaping on the analyzed modulation formats performance , the procedure to obtain the results presented in Fig. 4.a. is performed again but now using 2-poles BFs (the rest of the assumptions remains the same). The corresponding results obtained under this new situation are presented in Fig. 4.b.

From Fig. 4 it can be seen that the out-of-band sharpness of the transceivers frequency response can significantly affect the system performance (although the 3-dB bandwidth remains the same) according to every analyzed modulation format. For instance, let us consider the transmission of 25 Gb/s using 10G technology transceivers (case commonly referred as 25G/10G, shown in the blue colored area of Fig. 4). In this situation, if the transceiver response is modeled using a 1-pole BFs approach, PAM-2 outperforms the rest of the modulation formats in the complete blue region. However, if the filters response model changes from one to two poles BF, it is difficult to maintain the previous conclusion, as observed in Fig. 4.b. As another example, let us now consider the transmission of 50 Gb/s using 10G technology (case commonly referred as 50G/10G, shown in the yellow colored area of Fig. 4). From Fig. 4.a. we can observe that EDB, ODB and even PAM-4 (with a strong penalty) could be feasible alternatives for the 50G/10G transmission if a 1-pole filter case is considered. Anyway, if we again change the filter response to a 2-poles profile scenario, we can see from Fig. 4.b. that none modulation format are compliant in this 50G/10G scenario (maybe only EDB or ODB, with very strict filter requirements, so in practice developing custom 10G technology transceivers).

These two examples (among many) shows the performance dependency of the modulation formats versus the transceivers frequency response shape, not only in terms of the 3-dB bandwidth (as commonly done), but also considering the huge impact of the out-of-band filter sharpness. In the present contribution, the out-of-band filter shaping is characterized by means of the frequency at -20 dB attenuation: $f_{20dB}$ or $B_{20dB}$ parameter. Note that another reference attenuation value can be also chosen to measure the out-of-band bandwidth. The use of the -20 dB attenuation value was selected arbitrarily.

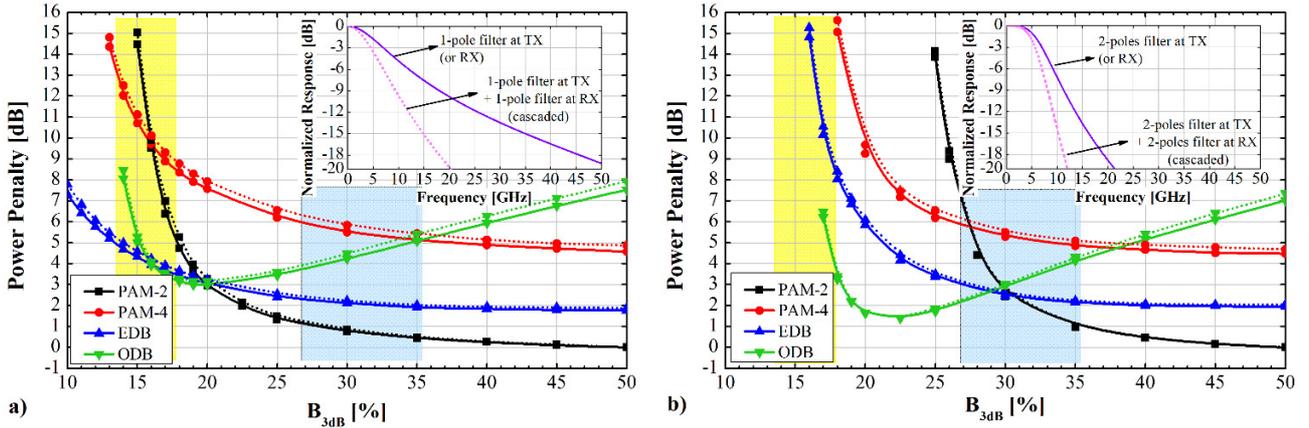

Fig. 4. Performance comparison among modulation formats in terms of Power Penalty with respect to S0, the PAM2 Sensitivity in the best condition (BtB unlimited bandwidth case),,S0 = -28.1 dBm for 25 Gb/s and S0 = -25.7 dBm for 50 Gb/s) as a function of the $B_{3dB}$ of the filters. a) 1-pole and b) 2-poles, BFs are employed (depicted in the inset). Solid lines for $R_b$ = 25 Gb/s, dotted lines for $R_b$ = 50 Gb/s. Colored regions: in blue (yellow): 10G technology to transmit 25 (50) Gb/s.

A first approach to start the out-of-band filter impact analysis consists of fixing the $B_{3dB}$ parameter (equal to 28%, a representative state-of-the-art case of the 25G/10G or 50G/25G transmission (Low cost transceivers already developed for 25 Gb/s transmission in data center, and used in the 50 Gb/s PON scenario, see Tables 1 and 2) and vary the number of poles of the BFs. As can be seen from Fig. 2.a., for Butterworth filter with a given $f_{3dB}$, increasing the number of poles, it corresponds to an increasing in the out of band filter sharpness, so a decreasing in its 20-dB bandwidth ($f_{20dB}$ or $B_{20dB}$). The performance of the four modulation formats in terms of sensitivity as a function of the BF number of poles (or the corresponding $B_{20dB}$ parameter) is shown in Fig. 5 (markers only), for 25 Gb/s transmission. Since the number of poles is an integer parameter, only certain discrete $B_{20dB}$ values can be evaluated. In order to overcome this $B_{20dB}$ granularity limitation, the use of the GFs model to emulate the transceiver frequency response is introduced. Under this GF approach, continuous curves of sensitivity versus $B_{20dB}$ can now be obtained, which are depicted in Fig. 5 (continuous curves without markers). Moreover, the comparison between the BF- and the GF-based results are useful to extract more information on the degree of tolerance of every modulation format when changing the transceivers frequency response shape (at least for two exemplary approaches).

A very close agreement between the BF- and the GF-based results in the case of EDB indicates a good resilience against changes in the filter shaping and lower power penalty with respect to the others modulation format. PAM-4 also exhibits a good tolerance when changing the filter profile for $B_{20dB} \geq 50\%$ but with very high power penalty. In contrast, PAM-2 and ODB have much less tolerance on changing the shape of the transceiver response. In the case of ODB, a very good agreement between the BF- and GF-based results is obtained for $B_{20dB} \leq 60\%$. However, for $B_{20dB} > 60\%$ values, the sensitivity results become completely different. The increased penalty shown with the GF approach is attributed to the reduced in-band power that this filter collects with respect to the BFs for higher values of B20dB (see Fig. 2). In the case of PAM-2 a good agreement is only found when the filter steepness is very low (around the 1-pole BF case). Please note that for PAM-2 a point between the 1- and 2-poles cases was plotted. This point was measured by setting a 1-pole filter at the TX and a 2-pole filter at the RX (which corresponds to a 3-pole filter when cascaded the TX and RX filters, emulating a symmetric 1.5-poles at the TX and RX situation). This is the only exceptional point that was evaluated using different characteristics at TX and RX filters in this contribution. However, following the same approach, the performance of the system using a 1-pole filter at TX and a 3-pole filter at RX was also tested for PAM-2 (corresponding to a 4-pole filter when cascading the TX and RX, equivalent to have 2-poles filter at both the TX and RX if a linear system is assumed). The sensitivity measured under this situation was the same as when having 2-pole filters at TX and RX, which shows that the system behavior is mostly linear, and the cascaded assumptions described in Section 3 can be considered accurate.

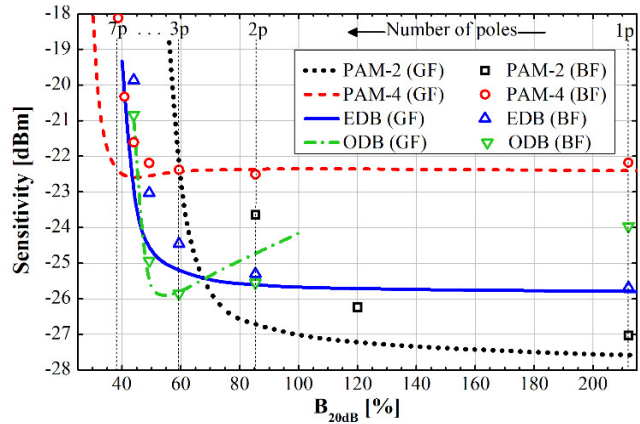

Fig. 5. Performance comparison in terms of B20dB using Butterworth (only marker curves) and Gaussian filters (only line curves). A 25 Gb/s transmission is considered. The B3dB is fixed to 28% (F3dB = 7 GHz for Rb = 25 Gb/s).

Regarding the modulation format tolerance against the variation of the $B_{20dB}$ parameter, again PAM-4 and EDB show the best degree of resilience, being PAM-4 the most robust format. By having a look on Tables 1 and 2 we can see that the $B_{20dB}$ parameter can vary from around 60 to even 140%, for different state-of-the-art devices. This fact highlight the need of considering the robustness of a modulation format against both in-band and out-of-band filter shaping variations as a relevant parameter.

The results shown in Fig. 5 were obtained for a fixed value of $B_{3dB} = 28\%$. For different $B_{3dB}$ values, the conclusions may change. In order to focus the filter shaping impact on the system performance, a wider analysis in which the joint variation of the $B_{3dB}$ and $B_{20dB}$ parameters have been performed. To this end, in Fig. 6 is depicted the power penalty of PAM-2 with respect to its better performance condition sensitivity S0 (S0 = -28.1 dBm for $R_b$ = 25 Gb/s, and S0 = -25.6 dBm for $R_b$ = 50 Gb/s) as a function of both $B_{3dB}$ and $B_{20dB}$ variables. The contour plots in solid lines correspond to the 25 Gb/s case, while the ones in dotted lines to the 50 Gb/s transmission. A very good agreement between the solid and dotted curves is again found. The real transceivers frequency response characteristics presented in Table 1 and Table 2 are also displayed in Fig. 6, in which any pair of $B_{3dB}$ and $B_{20dB}$ is considered as the coordinates of a point in the plane (indicating the operation regions of the current devices). Note the huge transceiver filter shaping impact on the performances: i.e in the case of using 25Gb/s transceivers for 50 Gb/s transmission (pink triangles), considering different real devices we can have negligible penalty lower than 0.2 dB, but in other cases large penalty until 10 dB. The same information is plotted in Fig. 7 for the others modulation formats, always evaluating the power penalty with respect to the same PAM-2 S0 sensitivity values of Fig. 6 of. Since the same good agreement between the 25 and 50 Gb/s results found for PAM-2 was also corroborated for the others modulation formats, only 50 Gb/s curves (representing both bit rate situations) are shown. This 3-dimensional power penalty representation as a function of the TX/RX frequency response characteristics shows the filtering impact on the system performance for every modulation format, and represents a good tool for analysis and design purposes.

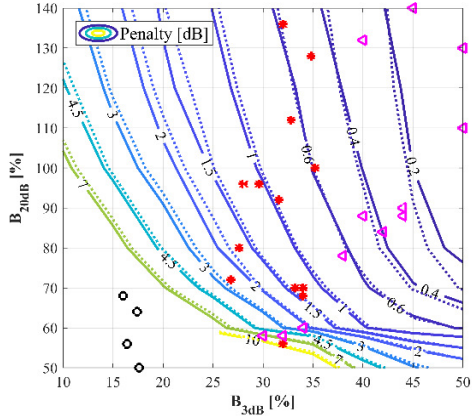

Fig. 6. PAM-2 power penalty with respect to the S obtained for PAM-2 in the best condition (BtB unlimited bandwidth case, S0 = -28.1 dBm for 25 Gb/s and S0 = -25.7 dBm for 50 Gb/s) as a function of the GFs $B_{3dB}$ and $B_{20dB}$ parameters. Red points: 25Gb/s using 10G technology transcheivers. Black circles: 50Gb/s using 10G technology transceivers. Pink triangles: 50Gb/s using 25G technology transceivers.

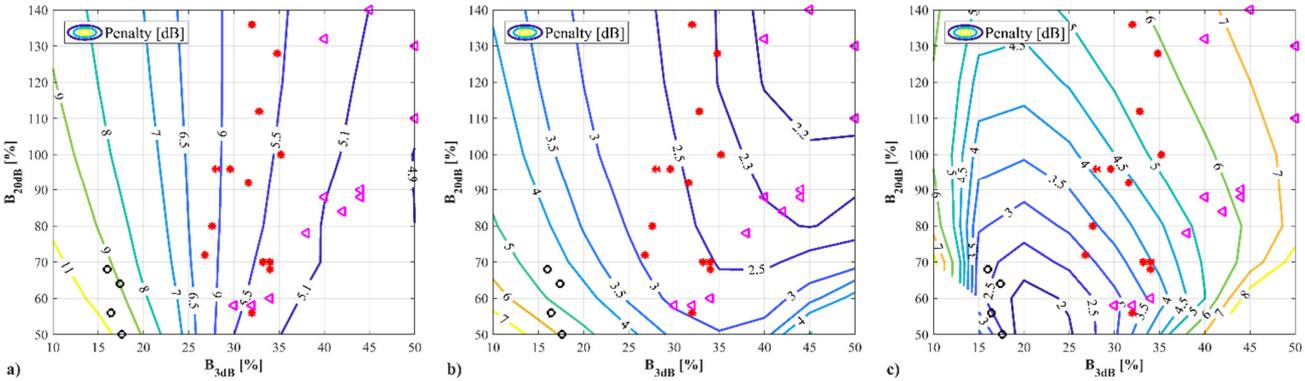

Fig. 7. PAM-4 (a), EDB (b), and ODB (c) power penalty with respect to the S obtained for PAM-2 in the BtB unlimited bandwidth case, S0 (S0 = -28.1 dBm for 25 Gb/s and S0 = -25.7 dBm for 50 Gb/s) as a function of $B_{3dB}$ and $B_{20dB}$ of the GFs. Red points: 25Gb/s using 10G technology. Black circles: 50Gb/s using 10G technology. Pink triangles: 50Gb/s using 25G technology.

By summarizing all the information provided by Figures 4 – 7, we can state the following conclusions with respect to the filter shape variation effect on the performance, regarding the BtB situation:

i) To transmit 25 Gb/s using 10G technology transceivers (25G/10G):
- PAM-4 exhibits the best tolerance against filter shaping variations. However, this format has a bigger penalty as compared to EDB or ODB in the 25G/10G region (see the red points of Figures 7.a., 7.b. and 7.c., and their associated power penalty).
- ODB exhibits large penalty variations as a function of the $B_{20dB}$ parameter. It is the only modulation format whose performance improves as the steepness of the filter increases. Therefore, we consider the use of this format convenient only in strong bandlimited situations.

- PAM-2 presents a strong filtering-dependent performance (see Fig. 6). For instance, by using two transceiver having the same $B_{3dB}$ = 30% but very different steepness, i. e. $B_{20dB}$ = 60 and 120% (not far from a real situation, see red points in Fig. 6), a very different penalty of >7 dB versus ~1 dB, respectively, can be obtained. Nevertheless, for systems with low bandwidth limitations, PAM-2 can achieve the best performance, i.e. the lowest penalty, being able to achieve <1 dB penalty (not achievable by any other format in any filtering situation).
- EDB shows a good tolerance against filtering, and a penalty between 2 – 3 dB in all the cases reported in Table 1. Accordingly, EDB is an interesting alternative in terms of resilience against filtering variations (with respect to both 3-dB bandwidth and steepness).

ii) To transmit 50 Gb/s using 10G technology (50G/10G)
- PAM-2 is not feasible (see the black circles of Fig.6).
- Neither PAM-4 or EDB seems to be feasible alternatives, since a large power penalty is achieved in this situation (see the black circles of Fig.7.a. and 7.b.).
- The only modulation format that may be used is ODB (see the black circles of Fig.7.c). However, in the $B_{3dB}$ < 15% range, the penalty of ODB rapidly increases even with small 3-dB bandwidth decreases. Therefore, the performance stability can be critical.

iii) To transmit 50 Gb/s using 25G technology (50G/25G):
- For PAM-2 the same conclusions as in case i) can be extrapolated. However, most of the 25G technology have more relaxed bandwidth limitations to transmit 50 Gb/s (see pink triangles in Fig. 6). Then, PAM-2 seems to be a good alternative to transmit 50G/25G.
- PAM-4 and ODB are not good candidates since a large power penalty (> 3 dB) arises in this situation.
- For EDB the same conclusions as in case i) can be extrapolated. However, in this 50G/25G case, EDB only outperform PAM-2 if transceivers with strong bandwidth limitations are employed.

Note that the previous conclusions can change when considering the joint effect of the dispersion and the bandwidth limitations in the following Section 5.

## 5. DISPERSION ANALYSIS

In Section 4, the effect of the transceivers filtering characteristics on the performance of four IM/DD modulation formats in a BtB scenario was analyzed. In this Section, the previous analysis is extended by also considering the presence of chromatic dispersion in the link.

As a first approach, the $B_{3dB}$ and $B_{20dB}$ parameters are fixed and the total dispersion of the link is varied in order to compute the sensitivity of the four modulation formats as a function of this last parameter. For the 25 Gb/s transmission, two representative {$B_{3dB}$, $B_{20dB}$} pairs were selected: P1={32%, 136%} and P2={32%, 56%} (cases 5 and 3 of Table 1, respectively), in order to compare the sensitivity versus dispersion curves obtained when the 3-dB bandwidth of the filters is the same, but the steepness is abruptly changed. The results are displayed in Fig.8.a. For the 50 Gb/s situation, the following {$B_{3dB}$, $B_{20dB}$} pairs were employed: P3={32%, 132%} and P4={32%, 58%} (case 5 of Table 2). The corresponding results are shown in Fig.8.b.

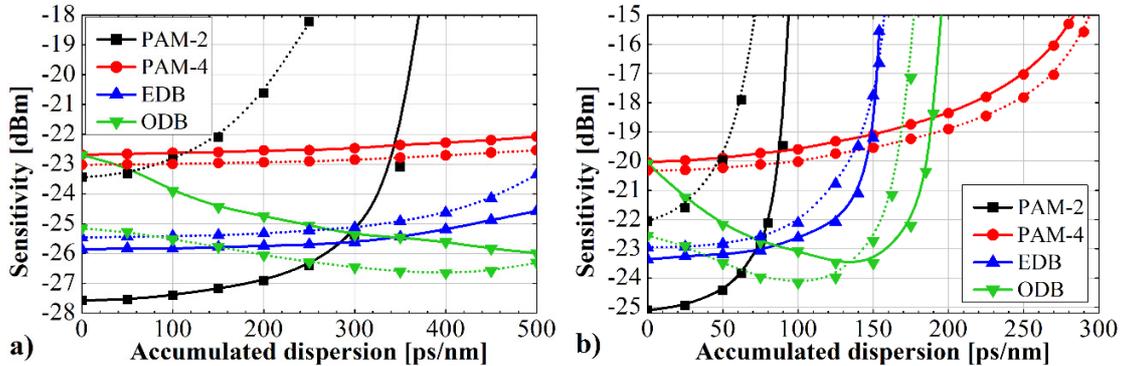

Fig. 8. Performance comparison among the four modulation formats in terms of sensitivity as a function of dispersion, a) $R_b$ = 25 Gb/s, solid: $B_{3dB}$ = 32%, $B_{20dB}$ = 136%. Dotted: $B_{3dB}$ = 32%, $B_{20dB}$ = 56%; b) $R_b$ = 50 Gb/s, solid: $B_{3dB}$ = 32%, $B_{20dB}$ = 132%, dotted: $B_{3dB}$ = 32 GHz, $B_{20dB}$ = 58%.

From Fig. 8.a. ($R_b$ = 25 Gb/s) we can observe that PAM-4 is the most robust format against the effect of both dispersion and filtering. EDB also exhibits a good tolerance against dispersion. Considering the transmission over CSMF with a standard PON fiber length of 20 km ($l$ = 20 km), EDB outperforms PAM-4 in O (~100 ps/nm), C (360 ps/nm) and L (~460 ps/nm) optical bands. Regarding PAM-2, as in BtB, in presence of dispersion its performance is also highly affected by the steepness of the out-of-band transceivers response. Its use in C- or L-band is not feasible for $l$ = 20 km. In O-band, its use seems to be strongly constrained to the use of technology with low steepness filtering characteristics. Under these conditions, its performance is the best among all modulation formats. ODB, on the other hand, is the only format in which the performance improves as both the dispersion and steepness of the filters increases (at least in O, C and L bands with $l$ = 20 km). In O-band ODB is outperformed by EDB, while in C- and L- bands seems to be the best alternative.

Regarding the 50 Gb/s transmission (see Fig. 8.b.) over 20-km of fiber, we observed that no modulation format can be used to operate the system in C-band or beyond. PAM-2 does not even work in O-band. The only feasible modulation formats (in O-band, $l$ = 20 km) are EDB, ODB and PAM-4, being the performance of PAM-4 surpassed by that of both EDB and ODB in the whole O-band.

Although PAM-4 has been found to be the most resilient format against dispersion and bandwidth limitations in all the analyzed conditions, due to its higher penalty as compared to EDB in all the practical scenarios, we do not consider it as a feasible alternative for the implementation of 25 Gb/s or 50 Gb/s PON systems over 20 km of fiber. Therefore, we do not analyze this modulation format in the rest of this Section. A similar consideration has been performed with respect to PAM-2 operating on C-band with Rb = 25 Gb/s, and O-band with Rb = 50 Gb/s. Therefore, besides the 25 Gb/s O-band case in which PAM-2 can still be considered a feasible format, the rest of this Section will be focus on the comparison between EDB and ODB formats.

Some preliminary results allow us to forecast the feasibility of PAM-2 and PAM-4 using pre-chirping in the transmission of 50 Gb/s in O-band and C-band (l = 20 km), respectively, at least under relaxed bandwidth limitations (1-pole BFs case). However, these alternatives need to be further explored under more strict filtering conditions, analysis that is out of the scope of the present contribution.

a.  25 Gb/s transmission results

In Figures 9 and 10 are shown, for O-band and C-band operation over 20-km of fiber, respectively, contour plots of the power penalty as a function of $B_{3dB}$ and $B_{20dB}$, for different modulation formats. The power penalty is evaluated in all cases with respect to S0 of PAM-2 in BtB conditions (S0 = -28.1 dBm for $R_b$ = 25 Gb/s).

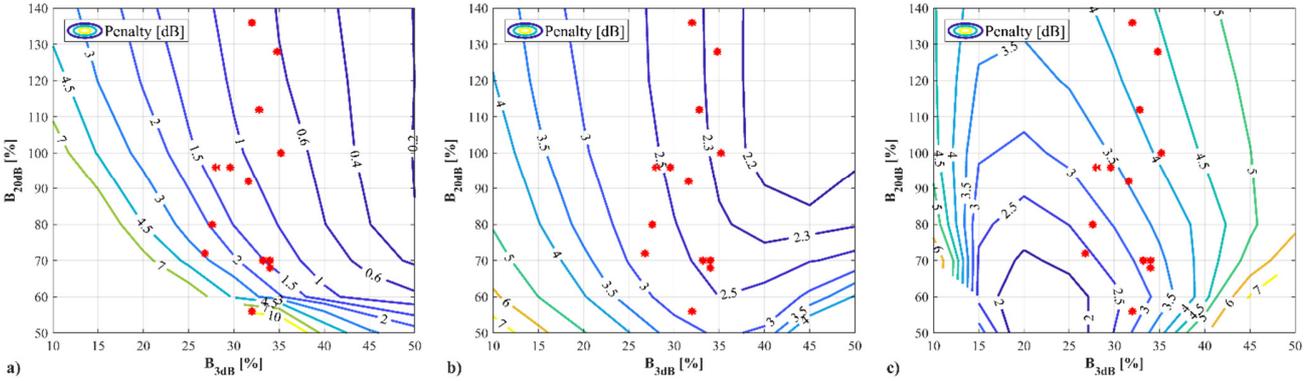

Fig. 9. PAM-2 (a), EDB (b), and ODB (c) power penalty with respect to the S obtained for PAM-2 in the BtB unlimited bandwidth case, S0 (S0 = -28.1 dBm) as a function of $B_{3dB}$ and $B_{20dB}$ of the GFs, for a 25 Gb/s 20-km O-band operation. Red points: 25 Gb/s using 10G technology.

Regarding O-band operation (Fig. 9), the three modulation formats exhibit similar results to the BtB situation (see Fig. 6 and 7), but having a small dispersion penalty in the case of PAM-2 and EDB. In contrast, ODB exhibits a performance improvement thanks to dispersion, which is consistent with the results presented in Fig. 8. By comparing the three modulation formats, we corroborate the strongest dependency of PAM-2 and ODB performance on the in-band and out-of-band transceiver response characteristics as compared to EDB. The power penalty of EDB varies between 2.2 – 3 dB when using state-of-the-art transceivers with different 3-dB bandwidth and out-of-band filtering steepness. In contrast, in the case of PAM-2, this penalty can vary from around 0.6 to 10 dB. ODB is an intermediate case (the penalty varies from 2.5 to 4.5 dB). Therefore, we consider EDB as the best alternative for 25 Gb/s 20-km O-band operation with respect to tolerance against frequency response variations. However, PAM-2 can be a good solution (lowest penalty) if (and only if) technology with proper filtering characteristics ($B_{3dB} \geq 30\%$ and $B_{20dB} \geq 70\%$) can be guaranteed.

With respect to C-band operation, we can see from the contour plots shown in Figures 10.a and 10.b, that the performance of ODB is further improved by dispersion while the opposite occurs for EDB. This situation tips the scales on favor of ODB in terms of lower range of power penalty achievable using state-of-the-art devices, i. e. from 1 to 3 dB, in contrast to 2.5 to 3.5 dB achieved with EDB. Anyway, EDB remains being the format with less performance variations as a function of filtering. It is important to note that, by using ODB and small bandlimited transceivers (i.e. $B_{3dB} \geq 35\%$ and $B_{20dB} \geq 100\%$) it is always possible to increase the performance of the system by intentionally adding an electrical filter at the TX or RX, in order to enforce a stronger bandlimited condition (close to the optimum point that can be seen in Fig. 10.b.) in which the achievable penalty could be around only 1 dB. This feature is not achievable using any other modulation format. Another positive (and unique) feature of ODB is that, as can be seen in Fig. 8 and it was aforementioned, the dispersion penalty decreases as increasing the fiber length, which can compensate, to some degree, the increase of fiber attenuation as the fiber length augments. In favor of EDB is the fact that it can be implemented with both direct modulation and external modulation (using an electro-absorption modulators or a MZM) approaches, while ODB needs a MZM as a must.

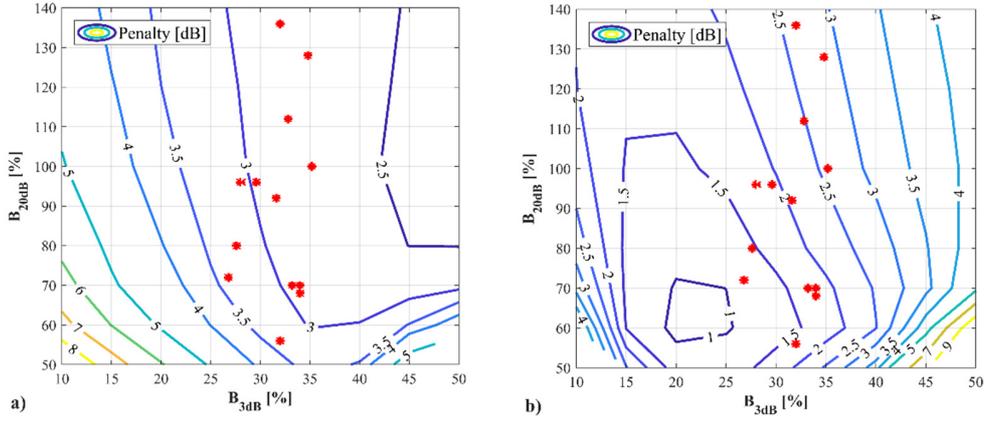

Fig. 10. EDB (a), and ODB (b) power penalty with respect to the S obtained for PAM-2 in the BtB unlimited bandwidth case, S0 (S0 = -28.1 dBm) as a function of $B_{3dB}$ and $B_{20dB}$ of the GFs, for a 25 Gb/s 20-km C-band operation. Red points: 25 Gb/s using 10G technology.

b. 50 Gb/s transmission results

As mentioned before, no modulation (without pre-chirping) is feasible for 20-km operation in C-band for a bit rate of 50 Gb/s. In addition, PAM-2 (without pre-chirping) cannot work either in O-band in this scenario, and PAM-4 is outperformed by both EDB and ODB in the whole O-band. In Fig. 11, contour plots of power penalty as a function of $B_{3dB}$ and $B_{20dB}$ for 50 Gb/s 20-km O-band operation using EDB (Fig. 11.a.) or ODB (Fig. 11.b.), are shown. The power penalty is referred to the 50 Gb/s PAM-2 S0 = -25.6 dBm. The contour plots displayed in Fig. 11 are very similar to those shown in Fig. 10 (for 25 Gb/s transmission in 20-km O-band). However, the regions of operation of 50G/10G and 50G/25G state-of-the-art devices (see the black circles and pink triangles of Fig. 11, respectively) are different of those of 25G/10G (see the red points of Fig. 10). Therefore, the conclusions between these cases may differ. First, ODB appears to be the only feasible format to transmit 50 Gb/s using 10G technology in O-band. However, we should take care of the filtering characteristics of the transceivers. In black circles of Fig. 11, only cases 2, 3, 10 and 11 of Table 1 are plotted. The rest of the cases has a $B_{20dB}$ <50%, which correspond to power penalties higher than 2.5 dB (not shown in Fig. 11.b.). Then, for correct operation, the $B_{20dB}$ of the TX and RX should be at least higher than 50%.

Regarding the transmission of 50 Gb/s using 25G technology, in both cases the maximum penalty shown in Fig. 11.a. (EDB) and Fig. 11.b. (ODB) is 4 dB. Again, EDB exhibits a more stable penalty varying between 3 – 4 dB (mostly around 3 dB), while ODB penalty variation is wider: from 1.5 to 4 dB. In terms of tolerance against filtering we should opt for EDB. However, as mentioned before, with the right technology characteristics or by enforcing the frequency response by means of an additional electrical filter, we could operate close to the optimal point of ODB and achieve a penalty as low as 1 dB.

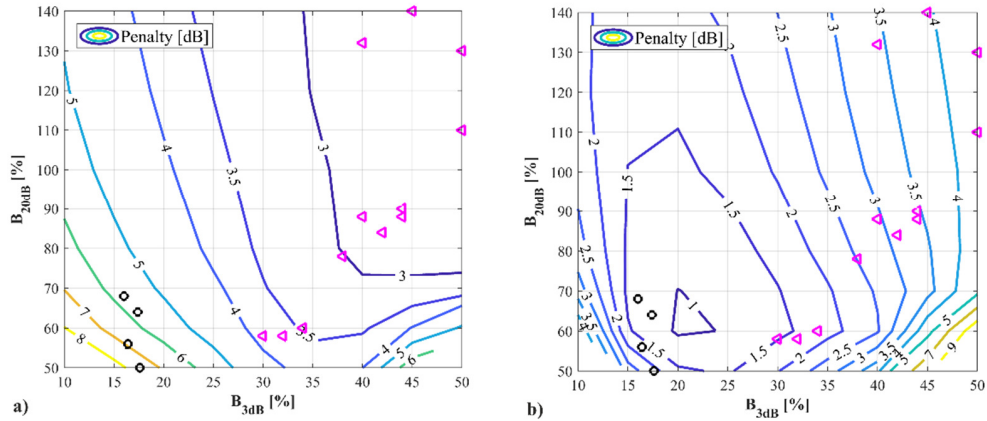

Fig. 11. EDB (a), and ODB (b) power penalty with respect to the S obtained for PAM-2 in the BtB unlimited bandwidth case, S0 (S0 = -25.6 dBm) as a function of $B_{3dB}$ and $B_{20dB}$ of the GFs, for a 50 Gb/s 20-km O-band operation. Black circles: 50 Gb/s using 10G technology. Pink triangles: 50 Gb/s using 25G technology.

## 6. CONCLUSIONS

In this contribution, we have demonstrated the strong impact on the system performance that the overall filter shaping electrical response of the transceivers has on the transmission of 25 and 50 Gb/s using currently available 10 and 25 Gb/s technology and adaptive equalization. By means of numerical simulations, we compared the achievable performance of PAM-2, PAM-4, EDB and ODB under different bandlimited and dispersive conditions. We introduced the 20-dB bandwidth parameter in order to quantify the impact of the out-of-band steepness of the transceivers response. We demonstrated that this parameter is as relevant as the 3-dB bandwidth when comparing the performance of different modulation formats.

We found that PAM-2 is the best performing format for the transmission of 25 Gb/s using 10G technology in 20-km O-band, but only if the use of transceivers with small bandwidth limitations can be guaranteed. Otherwise, the best alternative is EDB since it is more resilient to filter shaping variations than PAM-2, and outperforms the rest of the formats in terms of penalty.

On the transmission of 25 Gb/s using 10G technology in 20-km C-band, EDB exhibits a slightly higher maximum power penalty than ODB (4 dB vs 3.5 dB). The minimum achievable power penalty of ODB is 1.5 dB lower than that of EDB. However, the performance of EDB as a function of the 3-dB and 20-dB bandwidth is more stable (power penalty variations less than 1 dB) than in the case of ODB (power penalty variations of up to 2 dB).

Regarding the 50 Gb/s case, no modulation format works under 20-km of C-band operation. PAM-2 is not even feasible in O-band. ODB is the only format that can be used if using 10G technology (in 20-km O-band), but only if some filtering conditions can be guaranteed. If 25G technology is used and O-band is considered through 20-km of fiber, both EDB and ODB can work exhibiting a maximum penalty of 4 dB. Again, the penalty of ODB is the minimum under some filtering conditions, but EDB performance is more robust against filter shaping variations of the transceivers frequency response.

In all cases, PAM-4 is the most robust format against both dispersion and filtering conditions. However, due to its inherent higher power penalty with respect to the rest of the formats, PAM-4 is always outperformed by any of them in all the analyzed conditions.

## ACKNOWLEDGMENT

This work was supported by Telecom Italia under the grant "5G-PON" (2017). The work was also partially supported by the PoliTO PhotoNext center (www.photonext.polito.it).